\documentclass[%
 reprint,
 amsmath,amssymb,
 aps,
prl,
]{revtex4-2}

\usepackage{amssymb}
\usepackage{color}
\usepackage{ulem}
\usepackage{graphicx}
\usepackage{dcolumn}
\usepackage{bm}
\usepackage[toc,page]{appendix}
\usepackage{tikz}
\usepackage{amsmath}
\usepackage[english]{babel}
\usepackage[pdftex,plainpages=false,colorlinks=true,linkcolor=blue, citecolor=blue, urlcolor=blue]{hyperref}

\newcommand{\dx}{{d_{x^2-y^2}}}
\newcommand{\dz}{{d_{z^2}}}

\newcommand{\lno}{{\mathrm{La_3Ni_2O_7}}}

\begin{document}

\preprint{APS/123-QED}


\title{Electron Doping of $\lno$ Thin Films: Candidate Metal Dopants and Their Potential Impact on Superconductivity}

\author{Shi-Cong Mo}
\affiliation{ Guangdong Provincial Key Laboratory of Magnetoelectric Physics and Devices,  School of Physics, Sun Yat-sen University, Guangzhou, Guangdong 510275, China
}

\author{W\'ei W\'u}
\email[Corresponding author: ]{wuwei69@mail.sysu.edu.cn}
\affiliation{ Guangdong Provincial Key Laboratory of Magnetoelectric Physics and Devices,  School of Physics, Sun Yat-sen University, Guangzhou, Guangdong 510275, China
}

\date{\today}


\begin{abstract}
The bilayer Ruddlesden-Popper nickelate $\lno$ has emerged as a promising platform for
exploring and understanding high-temperature superconductivities. While existing doping studies have primarily
concentrated on hole doping  achieved through strontium  substitution or  oxygen content tuning, the electron-doped regime in this system remains largely unexplored. 
 In this work, we systematically investigate  possible  electron doping in $\lno$ thin films through
tetravalent element substitution,  employing first-principles density functional theory calculations.
 Our results suggest that $\mathrm{cerium}$ (Ce) doping  is inefficient in introducing electron carriers into the low-energy bands. In contrast, zirconium (Zr), hafnium (Hf), and thorium (Th) emerge as efficient electron donors.   We show that  Zr and Hf doping preferentially introduce electrons into  the $\dx$-derived bands,  while Th doping delivers more electrons into the $\dz$-derived bands. 
Electron dopings notably  augment the interlayer hopping $t_{\perp}$ between $d_{z^2}$ orbitals, which could  enhance superexchange coupling $J_{\perp}$   and  consequently promote an increase in superconducting $T_c$. 
 We evaluate the Coulomb interaction parameters using constrained random phase approximation.
Our results identify viable dopants for achieving electron-doped $\lno$, which not only diversifies the material family but also provides new  platforms for disentangling the origins of $\lno$ superconductivity.
\end{abstract}


\maketitle

\section{Introduction}
The discovery of superconductivity (SC) in nickelates~\cite{li2019_ni112, sun2023, mjiang_112,ding2023critical,karp20_112,lu2021magnetic}, particularly the bilayer Ruddlesden-Popper (RP) phase $\lno$ in both bulk~\cite{sun2023,li2026bulk,li_100gpa,chow2025bulk,jgchen24_nature} and thin film systems ~\cite{zxshen25_arpes,li2025enhanced,zychen25_film,osada2025strain,hao2025superconductivity,wang2026superconducting}, has drawn intense attention in condensed matter physics,  prompting a surge of experimental~\cite{zychen25_film,jgchen24_nature,yuanhq_24,jzhao24_4310,dlfeng24,meng2024density,lshu24_usr,xhchen25,jjzhang_25,zychen2025_odef,zhou2024revealing,yang_orbital24} and theoretical~\cite{zhluo23_prl,christiansson2023correlated,gsu24_tn,yyang_2compon23,fyang23_rpa,cjwu24_hund,zhluo24_qm,zheng2025s,Kuroki24,oh2023type,
lechermann23,wu24_charge,ku24,zhang24_dft,ryee2024quenched,jkun24,xtao24,jphu25,  zylu24_hund,liao2023electron,wang2025fermi,ryee2025optimal,maier2025,mo2025intertwined,chen2025electronic,zhong2026superexchanges,huang2025effective,chen2025charge} studies. With transition temperature ($T_c$) exceeding 80 K,  $\lno$ stands as a  rare analog to the iconic cuprate high-temperature superconductors. Similar to cuprates~\cite{armitage2010progress}, $\lno$ features quasi-two-dimensional $\mathrm{NiO_2}$ layers that host the dominant low-energy physics and exhibit strong electronic correlations. Crucial differences, however,  exist between the two families. In cuprates, strong Jahn–Teller effect and crystal field splitting lift the degeneracy of the $d$ orbitals, so that the electronic states near the Fermi surface arise primarily from the Cu-$3d_{x^2-y^2}$ orbital along with the associated O-$2p_x$ and -$2p_y$ orbitals. In contrast,   in  $\lno$ , both Ni-$3d_{x^2-y^2}$ and -$3d_{z^2}$ orbitals (the $e_g$ doublets) contribute to states near the Fermi level. The involvement of multiple $3d$-orbital-derived bands leads to rich and intricate physics, especially concerning the pairing mechanism, for which several competing scenarios have been proposed~\cite{zhluo23_prl,gsu24_tn,christiansson2023correlated,cjwu24_hund,fyang23_rpa,yyang_2compon23,ku24}.


Current theoretical studies have not reached consensus on the primary mechanism driving electron pairing in $\lno$~\cite{yyang_2compon23,cjwu24_hund,shao2025pairing,mo2025intertwined}. One central debate concerns the role of  $\gamma$ band, which originates from $\mathrm{Ni-}3d_{z^2}$ orbitals, in the pairing mechanism. While it has been proposed that the crossing of the Fermi level by the 
$\gamma-$ band is essential for superconductivity in $\lno$~\cite{zhluo23_prl,yyang_2compon23,sun2023,shen2026nodeless,gmzhang23_cpl,yyang_2compon23,berg08}.  other studies suggest that $s_\pm-$ wave pairing ~\cite{wen_gap,shen2026nodeless,sun2025gap} could still occur even when  $\gamma$ band lies below the Fermi level~\cite{yyang_2compon23,fyang23_rpa,cjwu24_hund,zhluo24_qm,zheng2025s}.  In the latter scenario, 
Hund's coupling $J_H$ between $e_g$ orbitals may play a role~\cite{cjwu24_hund,shao2025pairing}. Recently, using strong-coupling numerical calculations, we have proposed~\cite{mo2025intertwined}  that  in $\lno$,  when key parameters such as $\dz$ hole doping level $\delta_z$ are varied to drive a Lifshitz transition of  $\gamma$ band,  the nature of superconductivity may change as 
$\gamma-$ band crosses the Fermi level. In this picture, two competing pairing mechanisms~\cite{yyang_2compon23,cjwu24_hund}  could continuously evolve from one into the other upon $\dz$ doing.

In experiments,  introducing extra holes into the low-energy bands of  $\lno$ films can be achieved either by Sr doping or by reducing oxygen content~\cite{hao2025superconductivity,wang2026superconducting}. However, whether $\gamma-$ band can cross  Fermi level with varying Sr content remains unclear in experiments up to date~\cite{sun2025gap,zychen25_film}. Therefore, relying solely on hole doping may not be sufficient to  identify the role of the 
$\gamma-$ band in superconductivity. Electron doping, on the other hand, may offer an important alternative approach for this problem. As we will demonstrate in this work, electron doping can modify properties of $\lno$ in several ways, including changing the band structure,  varying filling levels of $\dx$/$\dz$ orbitals and importantly, shifting the $\gamma-$ band downward in energy. The dependence of SC on these properties may reveal crucial information about the pairing mechanism in $\lno$ that is inaccessible from hole-doped compounds alone, much like electron-doped  $\mathrm{Re_{2-x}Ce_{x}CuO_4}$ (Re = Pr, Nd, La)~\cite{armitage2010progress,skinta2002evidence} have contributed to the comprehension of cuprate superconductivity. Consequently, applying electron doping to $\lno$ is a natural and critical step for exploring $\lno$ superconductivity.

The primary goal of this work is to  investigate possible electron doping in $\lno$ thin films on substrates using \textit{ab initio} density functional theory (DFT) calculations.
 Thin-film $\lno$ samples offer notable advantages in experiments. For example, the dopant concentration can be continuously tuned~\cite{hao2025superconductivity,liu2026superconducting,wang2026superconducting},  the lattice constants can be modified by selecting different substrates, and stable phases can be obtained without applying high pressure~\cite{xu2024pressure,fan2026evolution}. These benefits make it easier to precisely control the doping level and to apply a wide range of experimental probes.

Our results demonstrate that substituting La with Zr, Hf, and Th in $\lno$ films   can effectively increase electron occupancy of the low-energy $e_g$ orbitals,
thereby achieving electron-doped $\lno$. In particular, Zr and Hf doping primarily inject electrons into the $\dx$-derived bands, shifting the near quarter-filled $\dx$ bands of the pristine compound toward half-filling, which  could enhance $\dx$-wave superconductivity within in-plane bands~\cite{zhluo24_qm}.
On the other hand, the interlayer hopping integral $t_{\perp}$ between $d_{z^2}$ orbitals is found to be significantly increased, which  may  also  leads  to a  higher $T_c$ of inter-layer $s\pm$-wave pairing in these electron-doped compounds. Using constrained random-phase approximation (cRPA), interacting parameters of $e_g$ orbitals of the electron-doped compounds are calculated. 
Finally, we investigate the influence of different substrates on the electron-doping behavior in $\lno$ thin films.

\section{Methods}
 Based on density functional theory, we performed first-principles calculations using the Vienna Ab initio Simulation Package (VASP)\cite{kresse1993ab,kresse1996efficient}. The projector augmented wave (PAW) pseudopotentials and the Perdew–Burke–Ernzerhof (PBE) exchange–correlation functional~\cite{perdew1996generalized} were employed, with a plane-wave cutoff energy of 600 eV. The Brillouin zones of the thin-film $\lno$ structures were sampled using $15 \times 15 \times 1$ k-point meshes, respectively. The convergence thresholds were set to $10^{-6}$ eV for electronic self-consistency and 1 meV/Å for ionic relaxation, ensuring reliable and well-converged results. To better account for the localized nature of Ni $d$-electrons, the DFT + $U$ method was applied with an effective $U$ parameter of 3.5 eV~\cite{yang_orbital24} for Ni-$3d$ orbitals. Constrained random-phase approximation (cRPA) method~\cite{aryasetiawan2004frequency} in VASP is used to estimate the effective screened interaction matrix of Ni-$3d$ electrons. 
 
\begin{figure*}[t!]
\centering 
\includegraphics[scale=0.4]{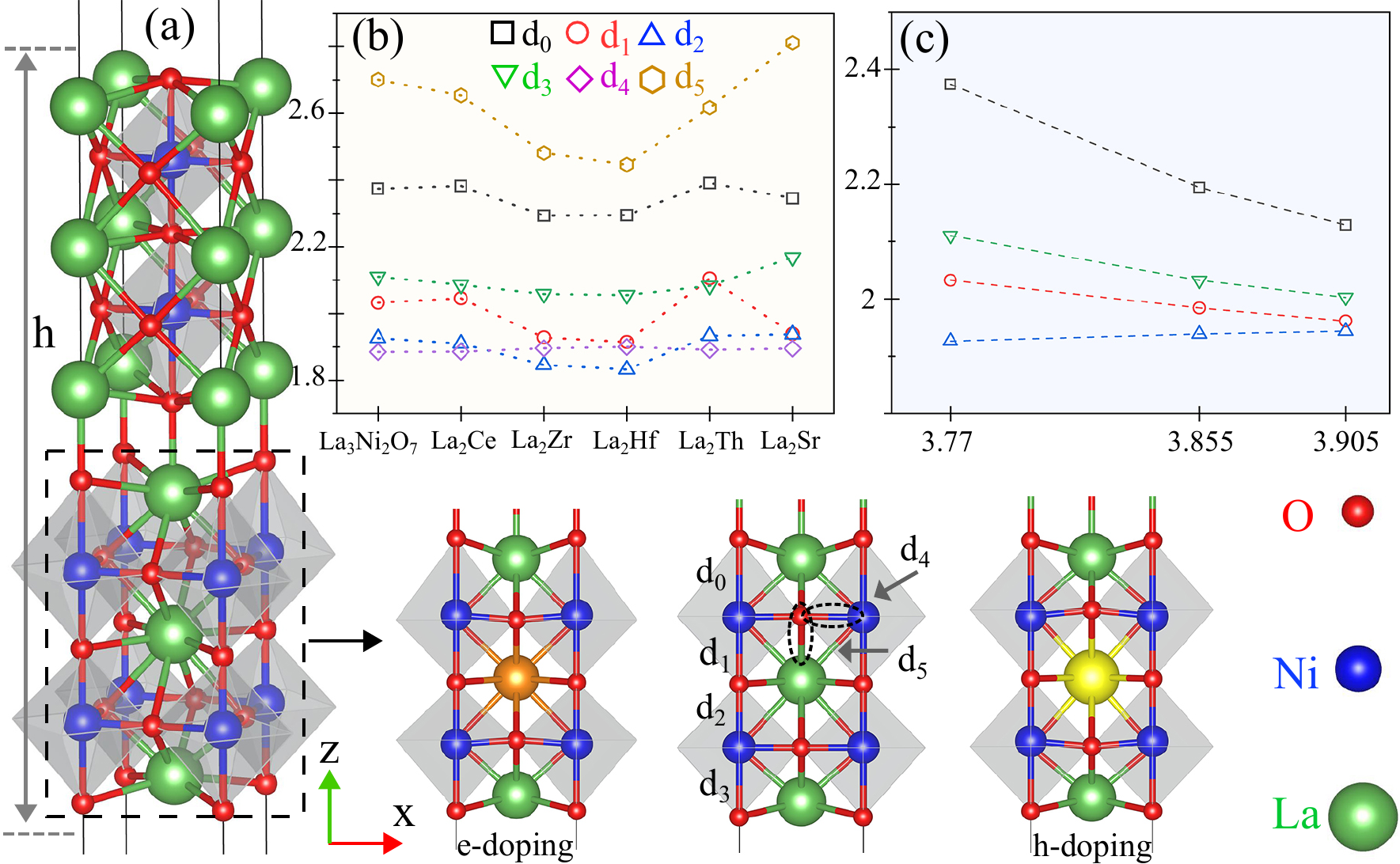}
\caption{(a): Structure of the 1-UC $\mathrm{La_{2-x}R_xNi_2O_7}$ thin film. Loci of electron/hole dopants are shown. Different chemical bonds are also labeled: $d_0–d_4$ denote different Ni–O bonds, and $d_5$ is the R–O bond. (b): Bond lengths of  thin-film $\lno$ and $\mathrm{La_2RNi_2O_7}$ (R = La, Ce, Zr, Hf, Th, and Sr).(c):  Bond lengths as functions of substrate in-plane lattices constant a.}
\label{fig:stru}
\end{figure*}

 For the thin-film systems, we  simulate one-unit-cell (1UC) thick $\mathrm{La_{3-x}R_xNi_2O_7}$ (R=Ce, Th, Zr, Hf) film on three different  representative substrates, $\mathrm{LaAlO_3}$~\cite{hhuang25_film}  with in-plane lattice constants of  a = 3.77 Å [ corresponding to a compressive  mismatch  strain of $\epsilon \sim –1.6\%$ ,  assuming  pseudo-tetragonal a = 3.833 Å of $\lno$ for reference ],  $\mathrm{NdGaO_3}$  with a = 3.855 Å ($\epsilon \sim +0.6\%$) ,  and $\mathrm{SrTiO_3}$with 3.905 Å ($\epsilon \sim +1.9\%$ )~\cite{osada2025strain}.  As we will see, our main conclusions are qualitatively the same across these substrates.  We take the inner-layer La site as the R- substitution position~\cite{shi2025theoretical}, as shown in Fig.~\ref{fig:stru}. A vacuum layer of 18 Å was introduced in the system to prevent spurious interactions between periodic images. We use $\mathrm{La_2SrNi_2O_7}$ as a reference hole-doped system for comparison with electron doping~\cite{shi2025theoretical,hao2025superconductivity,wang2026superconducting}.

\begin{figure}[t!]
\includegraphics[scale=0.38]{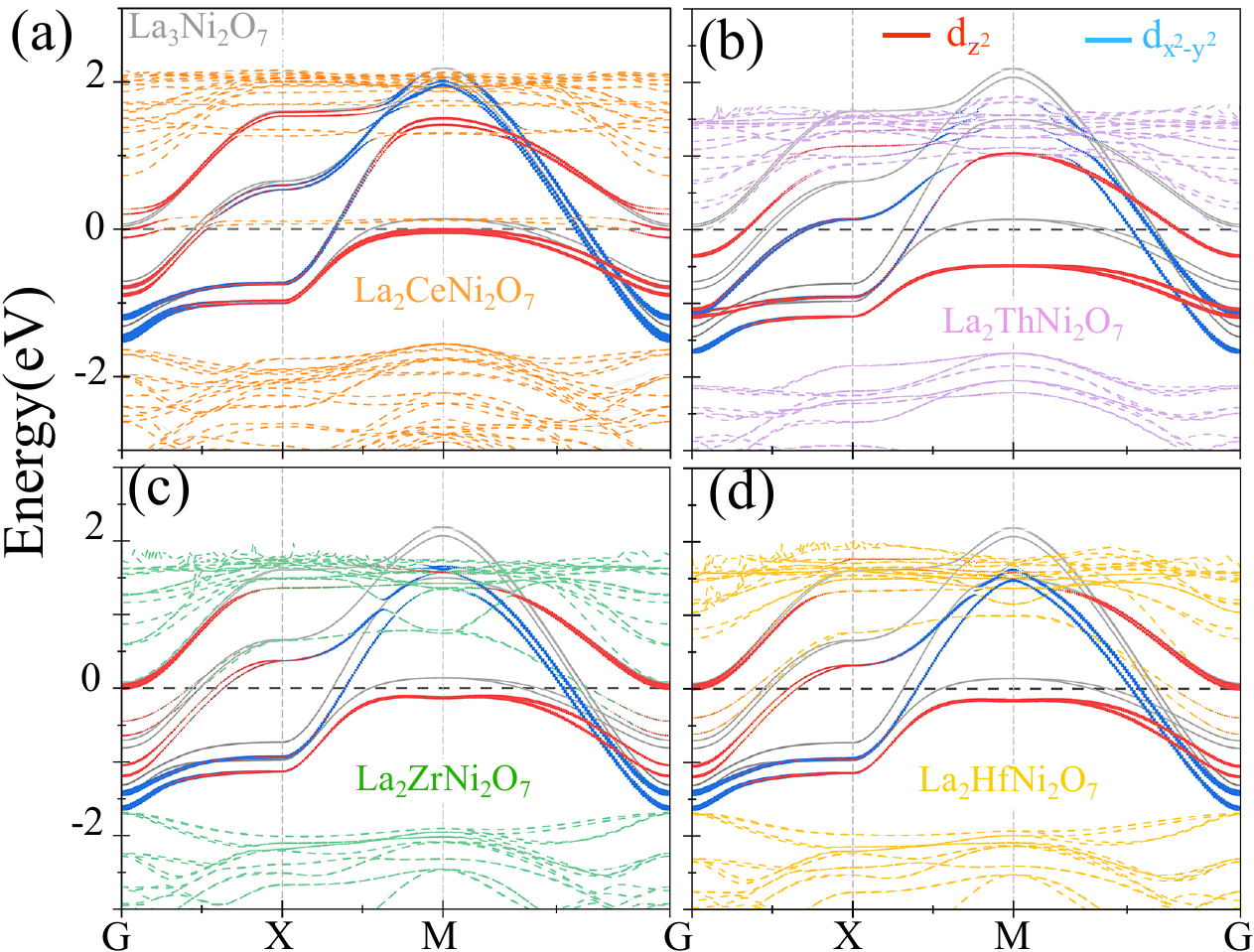}
\caption{Energy bands  of  1-UC $\mathrm{La_{2-x}R_xNi_2O_7}$ thin films. (a) $\mathrm{La_2CeNi_2O_7}$, (b) $\mathrm{La_2ThNi_2O_7}$, (c) $\mathrm{La_2ZrNi_2O_7}$, (d) $\mathrm{La_2HfNi_2O_7}$.  The $\dz$ and $\dx$ orbitals weights in  low-energy bands
are denoted by  red (dark) and blue (light)  dots respectively.  Low-energy bands  of  pristine $\lno$ film on substrate are shown by gray lines for reference. Here substrate with  $a=3.905$ Å  is assumed. For $\mathrm{La_2CeNi_2O_7}$ DFT+$U$ is also applied to the $4f-$ orbitals of Ce. See  also Appendix.}
\label{fig:band}
\end{figure}

\section{Result}
\subsection{Electronic structures}  
In cuprates,  substituting La by Ce can introduce mobile electrons into  the  $\mathrm{CuO_2}$  planes, enabling electron doping. Surprisingly, however, we find that the same substitution is substantially inefficient in $\lno$. This is demonstrated  in Fig.~\ref{fig:band}a, which compares the band structures of pristine $\lno$ and $\mathrm{La_2CeNi_2O_7}$.  One can see that even at a  Ce concentration level as high as $x$ = 1, the shifting of the low-energy bands~\cite{zhluo23_prl,yyang_2compon23,cjwu24_hund,nie2026superconductivity} of  $\mathrm{La_{2-x}Ce_xNi_2O_7}$  (highlighted)  relative to those of the pristine $\lno$ (gray lines) is very mild.  We also proceeded to  first  replace La with Nd ($\mathrm{Nd_3Ni_2O_7}$), then substitute Nd with Ce  ($\mathrm{Nd_{2-x}Ce_xNi_2O_7}$), but the electron doping effect in the low-energy bands remains weak (see Appendix). This behavior also differs from that in cuprate superconductors, where electron doping can be achieved by substituting $\mathrm{Nd^{3+}}$ with $\mathrm{Ce^{4+}}$~\cite{da2015charge}. 

We explore other tetravalent metal elements and  identify several more potent electron dopants: $\mathrm{La_{2-x}Th_xNi_2O_7}$, $\mathrm{La_{2-x}Zr_xNi_2O_7}$, and $\mathrm{La_{2-x}Hf_xNi_2O_7}$ (Fig.~\ref{fig:band}b–d). In all three compounds,   $\dz$ and $\dx$ bands near the Fermi level  are significantly  shifted downward in energy upon doping. Zr and Hf, belonging to the same group, produce similar electronic structures in the  energy window $\sim$ (–2 eV, +2 eV): the top of the $\dz$ bonding band at the M point drops to about $\sim –0.3$ eV, while the tops of the antibonding bands are strongly modified by electron hybridizations from other orbitals.  For Th doping, the $\dz$ bonding band top at M point descends to an even lower energy of $\sim –0.6$ eV, indicating stronger $\dz$ electron doping than in  cases of Zr and Hf.

Band structures of  doped $\lno$  on other substrates  can be found in the Appendix (Figs.~\ref{fig:band-strain} and \ref{fig:band-3.77}). A discussion of DFT treatment of the Ce $4f$-electron  and the lattice-constant dependence of Ce doping, including a comparison with cuprates, is provided in the Appendix (Figs.~\ref{fig:uf}, \ref{fig:strain-f}, and \ref{fig:cu-dope}).

\subsection{Structures}  
To gain in-depth understanding of the doping effects,  we now examine the structural changes upon doping. In Fig.~\ref{fig:stru},  lengths of different Ni–O bonds  [labeled as $d_0$ - $d_3$ (vertical), and $d_4$ (in plane) ] and the La–O bond ( $d_5$) are investigated.
As one sees in Fig.~\ref{fig:stru}b, introducing $\mathrm{Th^{4+}}$ causes elongated  Ni–O bonds $d_0$, $d_1$,  a tendency
just to the opposite of $\mathrm{Sr^{2+}}$  doping.  This is because electron doping shifts the average valence of Ni from $\mathrm{Ni^{2.5+}}$ toward $\mathrm{Ni^{2+}}$. Given a larger ionic radius of $\mathrm{Ni^{2+}}$, the expansion of the electron cloud leads to the elongation of the vertical Ni–O bonds. Conversely, hole doping in the $\mathrm{La_{2-x}Sr_xNi_2O_7}$ case drives $\mathrm{Ni^{2.5+}}$ to $ \sim \mathrm{Ni^{3+}}$,  resulting in the shortening of  Ni–O bonds.

For $\mathrm{Zr^{4+}}$ and $\mathrm{Hf^{4+}}$ doping,  Fig.~\ref{fig:stru}b shows that, however,  all three vertical  Ni–O bonds  $d_0$ to $d_2$  shrink, contrasting $\mathrm{Th^{4+}}$ doping. This phenomena arises because the ionic radii of $\mathrm{Zr^{4+}}$ (0.89 Å) and $\mathrm{Hf^{4+}}$(0.88 Å) are much smaller than that of $\mathrm{La^{3+}}$(1.16 Å). The strong lattice contraction effects caused by the excessively small ionic radius of R elements outweigh the electron‑cloud expansion effect stemming from Ni ion. On the other hand, the variation of La–O bond ( $d_5$) in different compounds generally reflects the ionic  radius of dopants relative to $\mathrm{La^{3+}}$ (1.16 Å). From $\mathrm{Hf^{4+}}$(0.88 Å) to   $\mathrm{Zr^{4+}}$ (0.89 Å) , $\mathrm{Th^{4+}}$(1.05 Å),  and to  $\mathrm{Sr^{2+}}$(1.26 Å, hole doping), a smaller ionic radius of R elements results in a shorter $d_5$ distance. Finally, the Ni–O bonds $d_3$ (near the substrate) and $d_4$ (horizontal) remain nearly unchanged upon doping. This is expected as they are more constrained by the fixed lattice constant of the substrate.
For $\mathrm{La_{3-x}Ce_xNi_2O_7}$ (Fig.~\ref{fig:band}a), we note that as shown in Fig.~\ref{fig:stru}b,  the Ni–O bonds and lattice structure of $\lno$ are almost not changed after Ce-doping. Suggesting Ce  retains largely a $\mathrm{Ce^{3+}}$ valence state, thus
 achieves only weak electron‑doping. Indeed, $\mathrm{Ce^{3+}}$ state has an ionic radius of 1.143 Å that being close to $\mathrm{La^{3+}}$, whereas $\mathrm{Ce^{4+}}$ has an much smaller ionic radius of 0.97 Å.
 
Fig.~\ref{fig:stru}c shows the variation of vertical Ni-O bond lengths $d_0$ to $d_3$ of pristine $\lno$  on different substrates ($\mathrm{LaAlO_3}$, 3.77 Å; $\mathrm{NdGaO_3}$, 3.855 Å; $\mathrm{SrTiO_3}$, 3.905 Å)~\cite{hhuang25_film,osada2025strain}. It can be seen that the vertical Ni-O bonds, except for $d_2$ that remains nearly constant,  generally decrease with increasing  in‑plane lattice constant. Doped compounds exhibit similar trend upon lattice constant variation (result not shown).

\begin{figure}[t!]
\centering 
\includegraphics[scale=0.275]{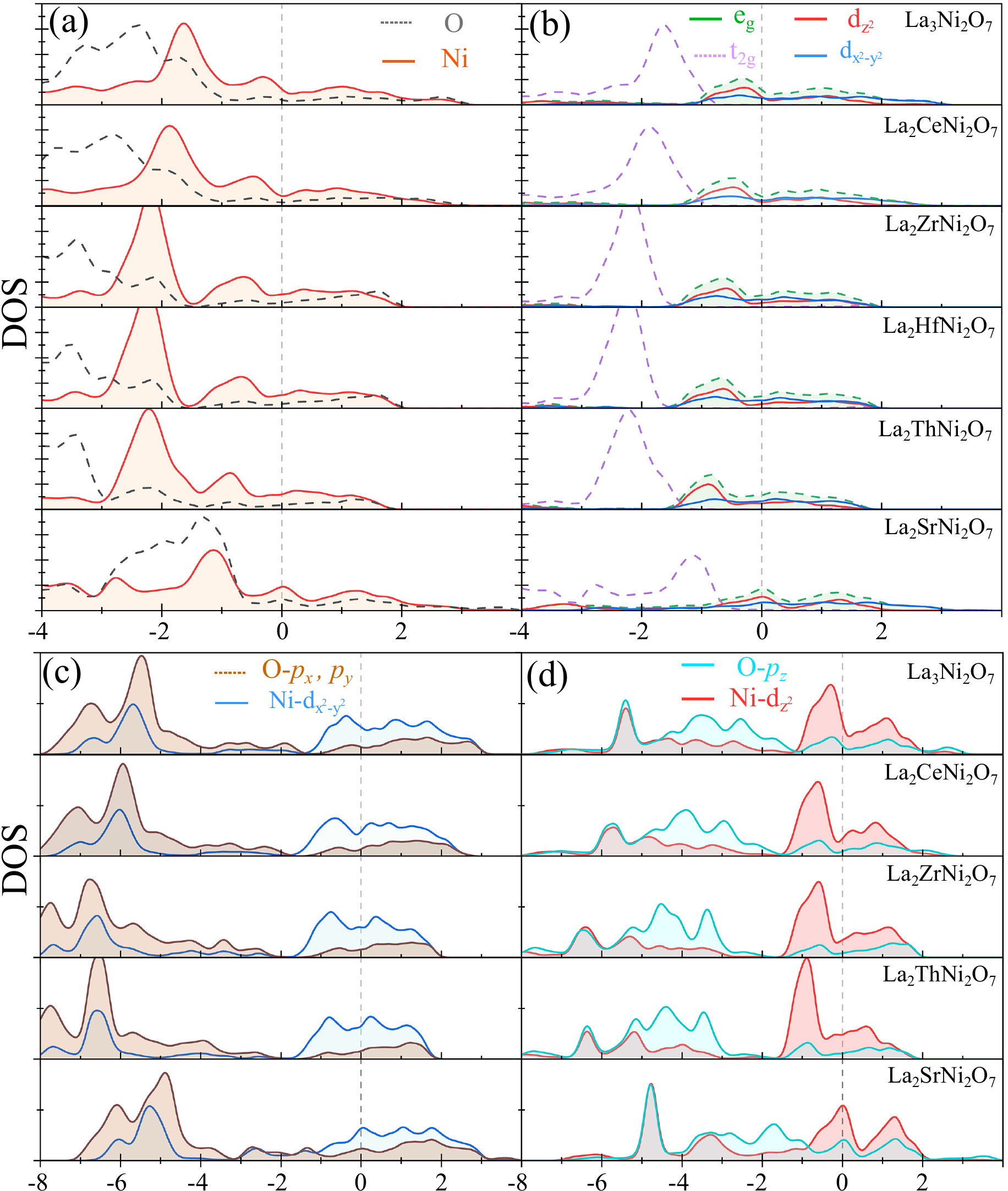}
\caption{DOS of  thin-film $\mathrm{La_2RNi_2O_7}$. In (c) and (d), DOS of in-plane Ni-$\dz$ orbital and its associated O-$p_x$ and -$p_y$ orbitals,  and the out-of-plane Ni $\dz$ orbital and its associated O-$p_z$ orbital are separately shown.}  
\label{fig:dos}
\end{figure}


\subsection{Orbital occupancy and charge transfer} 

We now perform density of states (DOS) calculations and electron counting to understand how doped electron carriers enters  the system.
 In Fig.~\ref{fig:dos}a, and  Fig.~\ref{fig:dos}b,  Ni exhibits a feature of two major  bumps  in low-energy DOS  arising from   $t_{2g}$ and $e_g$ orbitals, respectively. For electron dopants (Zr, Hf, Th), both features shift downward in energy as comparing the un-doped $\lno$.
Figure~\ref{fig:dos}b resolves the shifting of different $d$ sub-shells in different compounds, which clearly shows that Zr, Hf, and especially Th doping can shift both the $\dx$ and $\dz$ orbitals downward, confirming that doping increase the electron occupation of the active $e_g$ orbitals~\cite{shi2025effect}. Interestingly, here we can see that, the $t_{2g}$/$e_g$ splitting (i.e., the peak separation) are widen as substituting La with electron-dopants. This is  because smaller ionic radius of dopants cause lattice contraction, which enhances the crystal field strength.

In cuprates, according to the well-known Zhang–Rice singlet (ZRS) theory~\cite{zhang1988effective},  carriers in ligand orbitals can strongly hybridize with those on transition metal ions~\cite{kowalski2021oxygen}. Integrating out the ligand $p$ orbitals can lead to superexchange coupling between neighboring transition metal $d$ orbitals.  This  superexchange coupling dominates the low-energy physics and act as the main driving force for superconductivity~\cite{wu25interplay}. The ZRS theory has been widely used to explain various phenomena in cuprates. For $\lno$ , numerical and experimental studies~\cite{wu24_charge,dong2024visualization} have shown that ZRS could also be applicable for describing its low-energy physics, particularly those involving  $\dz$ orbitals. 

In ZRS theory,  the combined carriers in oxygen $p$ orbitals and transition metal $d$ orbitals form a low-energy spin-singlet band, known as the Zhang–Rice singlet band (ZRSB)~\cite{zhang1988effective}. To consider the effective carrier content  meaningful for low-energy physics, carriers in Ni-$3d$ orbitals  and the hybridizing O-$2p$ orbitals  thereby should be counted together. To this end, we plot in Fig.~\ref{fig:dos}c the DOS of the Ni-$\dx$ orbital together with its hybridizing in-plane O-$p_x$ and O-$p_y$ orbitals ( in-plane orbitals ),  and in Fig.~\ref{fig:dos}d   DOS of the $\dz$ orbital together with the out-of-plane O-$p_z$ orbital ( out-of-plane orbitals). From these plots, we see that as   DOS weight of  $3d$ orbitals is pushed to lower energies,  $2p$ bands are  driven to lower energies upon doping as well. Hence, both the $3d$ and $2p$ orbitals are electron-doped when Zr/Hf/Th dopants are introduced. For comparison, results of hole-doped $\mathrm{La_2SrNi_2O_7}$ are also shown in Fig.~\ref{fig:dos}, which generally exhibit trends opposite to those of electron doped cases. The $\mathrm{La_2CeNi_2O_7}$ case with weak doping, on the other hand, exhibits a modest downward shift of the DOS.

\begin{table*}
  \caption{Electron counts for Ni-$d$ orbitals, $t_{2g}$ orbitals, $e_g$ orbitals, as well as for the in-plane and out-of-plane components with a=3.77 Å.}
\setlength{\tabcolsep}{8pt}
\setlength{\extrarowheight}{3pt}
\label{tab:label}
  \centering
  \begin{tabular}{cccccccccc}
    \hline
      & Ni-$d$ & $t_{2g}$ & $\dz$ & $\dx$ & In & Out & $\Delta$In & $\Delta$Out \\
\hline
    $\lno$ & 8.360 & 5.826 & 1.412 & 1.123 & 3.231 & 3.005 & -  & - \\
    \hline
    $\mathrm{La_{2.5}Ce_{0.5}Ni_2O_7}$ & 8.310 & 5.803 & 1.411 & 1.097 & 3.221 & 3.012 & -1.0\% & 0.7\% \\
    $\mathrm{La_2CeNi_2O_7}$ & 8.424 & 5.808 & 1.454 & 1.162 & 3.298 & 3.063 & 6.7\% & 5.8\% \\
    \hline
    $\mathrm{La_{2.75}Zr_{0.25}Ni_2O_7}$ & 8.383 & 5.799 & 1.449 & 1.135 & 3.270 & 3.053 & 3.3\% & 4.8\% \\
    $\mathrm{La_{2.5}Zr_{0.5}Ni_2O_7}$ & 8.411 & 5.775 & 1.446 & 1.190 & 3.354 & 3.063 & 12.3\% & 5.8\% \\
    $\mathrm{La_2ZrNi_2O_7}$ & 8.448 & 5.735 & 1.453 & 1.260 & 3.464 & 3.083 & 23.3\% & 7.8\% \\
    \hline
    $\mathrm{La_{2.75}Hf_{0.25}Ni_2O_7}$ & 8.387 & 5.806 & 1.449 & 1.132 & 3.266 & 3.054 & 3.5\% & 4.9\% \\
    $\mathrm{La_{2.5}Hf_{0.5}Ni_2O_7}$ & 8.424 & 5.777 & 1.451 & 1.197 & 3.366 & 3.068 & 13.5\% & 6.3\% \\
    $\mathrm{La_2HfNi_2O_7}$ & 8.469 & 5.731 & 1.459 & 1.279 & 3.492 & 3.088 & 26.1\% & 8.3\% \\
    \hline
    $\mathrm{La_{2.75}Th_{0.25}Ni_2O_7}$ & 8.392 & 5.808 & 1.449 & 1.135 & 3.266 & 3.058 & 3.5\% & 5.3\% \\
    $\mathrm{La_{2.5}Th_{0.5}Ni_2O_7}$ & 8.416 & 5.785 & 1.480 & 1.151 & 3.305 & 3.102 & 7.4\% & 9.7\% \\
    $\mathrm{La_2ThNi_2O_7}$ & 8.476 & 5.742 & 1.537 & 1.197 & 3.390 & 3.183 & 15.9\% & 17.8\% \\
    \hline
    $\mathrm{La_{2.75}Sr_{0.25}Ni_2O_7}$ & 8.324 & 5.832 & 1.377 & 1.115 & 3.201 & 2.951 & -3.0\% & -5.4\% \\
    $\mathrm{La_{2.5}Sr_{0.5}Ni_2O_7}$ & 8.291 & 5.829 & 1.364 & 1.098 & 3.167 & 2.920 & -6.4\% & -8.5\% \\
    $\mathrm{La_2SrNi_2O_7}$ & 8.251 & 5.842 & 1.311 & 1.098 & 3.127 & 2.828 & -10.4\% & -17.7\% \\
    \hline
    $\Delta$In=($\mathrm{In_R}$ - $\mathrm{In_{La}}$) $\times$ 100\% \\
  \end{tabular}
\end{table*}

Integrating DOS weight  up to  the Fermi level yields the electron filling numbers for  different $\mathrm{La_{3-x}R_xNi_2O_7}$ compounds, as  summarized in Table~\ref{tab:label}. Taking R=Th as an example, the compositions $\mathrm{La_{2.75}Th_{0.25}Ni_2O_7}$, $\mathrm{La_{2.5}Th_{0.5}Ni_2O_7}$, and $\mathrm{La_{2}ThNi_2O_7}$ correspond to  in-plane (out-of-plane) orbital doping levels of 3.5\% (5.3\%), 7.4\% (9.7\%), and 15.9\% (17.8\%), respectively. It is worth noting that, in an effective low-energy model where the O-$p$ orbitals are integrated out, the out-of-plane orbitals map onto the $\dz$-derived bands, whereas the in-plane orbitals correspond to the $\dx$-derived bands. The electron/hole doping levels presented in Table~\ref{tab:label} thus serve as essential input for constructing effective models with properly assigned carrier densities per band. 

From Table~\ref{tab:label}, one can observe that or all three electron dopants, the actual doping levels are considerably lower than their nominal values. For instance, at $x=0.5$, the total doping 
levels are respectively 18.1\%, 19.8\%, and 17.1\% for Zr, Hf, Th, in contrast to the nominal 25\% doping.
More importantly, the relative ratio of carriers doped into in-plane versus out-of-plane orbitals (i.e., $\dx$- vs. $\dz$-derived bands) varies with the dopant species. For Th doping, more electrons enter  $\dz$-derived bands, whereas for Zr and Hf, more electrons move to  $\dx$-derived bands at large doping levels. 
In the latter case, the added electrons drive the near-quarter-filled $\dx$ bands of the pristine compound closer to half-filling, thereby enhancing in-plane antiferromagnetic correlations. This enhancement may facilitate the emergence of in-plane $\dx$-wave superconductivity in $\lno$~\cite{zhluo24_qm}. Thus, upon electron doping, one might anticipate a transition in the pairing symmetry from an interlayer $s\pm$ wave to an intralayer $\dx$ wave, or possibly a hybrid state combining both.

 

\begin{figure}[t!]
\includegraphics[scale=0.4]{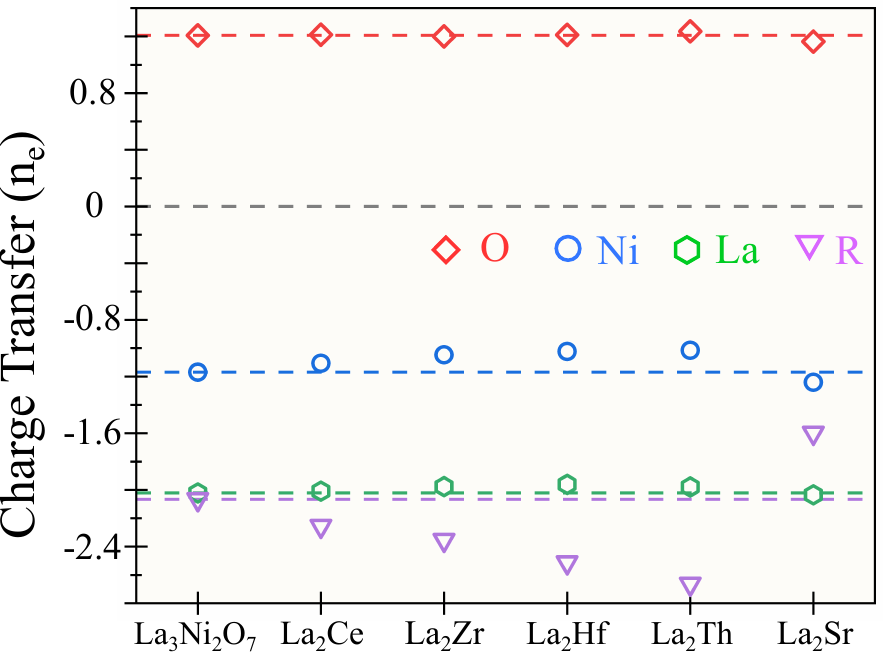}
\caption{ Average charge transfer between O, Ni, La, and R (La, Zr, Hf, Th, Sr) elements.  Here, O gains charge, while the other atoms lose charge. Dotted line denotes result of  $\lno$  for reference. }  
\label{fig:charge}
\end{figure}

We next perform Bader charge analysis~\cite{henkelman2006fast} to evaluate the charge transfer induced by doping. As shown in Fig.~\ref{fig:charge}, compared to La in pristine  $\lno$, electron-doping (triangles) enhance the negative charge donated to the system from dopant element R, whereas hole doping suppresses it. To preserve overall charge neutrality, the Ni valence correspondingly decreases for electron doping and increases for hole doping. In contrast, the oxygen sites and the unsubstituted La sites maintain relatively stable valences, with Bader charge transfer values of approximately 1.1–1.25 
$|e|$ and about 2.0 $|e|$, respectively, across the various compositions.

\subsection{Tight-binding parameters} 
In addition to the change of carriers, electron doping can also alert energy bands that may  significantly affect the superconducting properties. This is especially true, when the effective  $\dz - \dz$ interlayer hopping integral $t_{z\perp}$ is changed.  The magnitude of $t_{z\perp}$ can directly influence the strength of magnetic correlations that serve as the driving force for superconductivity~\cite{tian2024correlation,yi2025unifying}.  Indeed, in the superexchange scenario, the inter-layer magnetic coupling $J_{\perp}$ scales like  $J_{\perp} \sim t_{pd}^4/(U+\Delta_{pd})^3$~\cite{wu24_charge}, where $t_{pd}$ is inter-layer $d-p$ hybridization that determines  $t_{z \perp}$.

\begin{figure}[t!]
\includegraphics[scale=0.35]{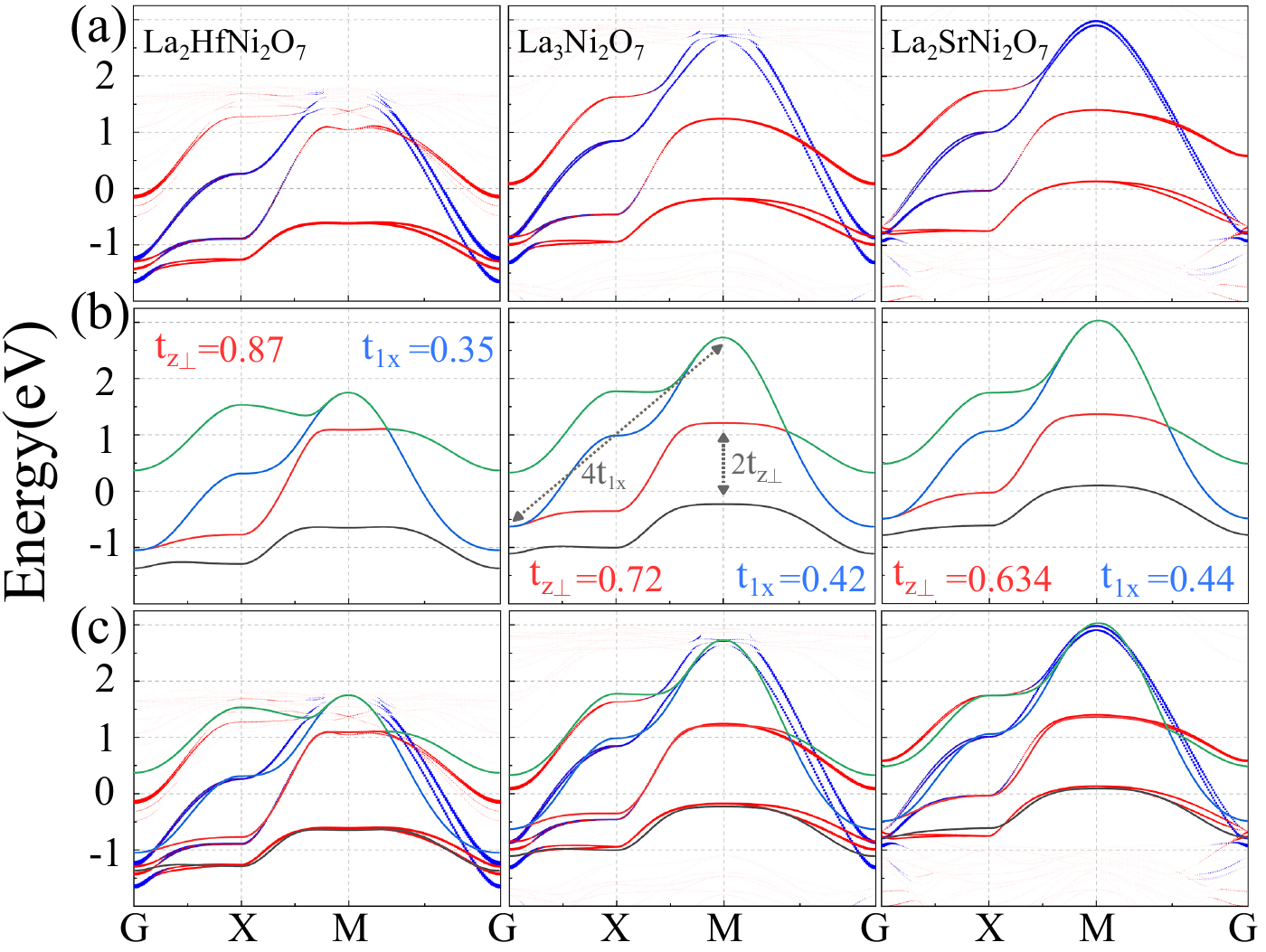}
\caption{ (a) Band structure of $\mathrm{La_{2}HfNi_2O_7}$, $\lno$, and $\mathrm{La_2SrNi_2O_7}$ calculated by first-principles. (b) Energy bands fitted by the double-layer Hubbard model. (c) Superposition of the energy bands in (a) and (b).
}
\label{fig:sr}
\end{figure}

We perform Wannier90~\cite{mostofi2014updated} fitting of the DFT band structures of $\mathrm{La_2HfNi_2O_7}$, $\lno$, and $\mathrm{La_2SrNi_2O_7}$ on the 3.77Å substrate to obtain typical tight-binding parameters. In particular, the values of inter-layer hopping $t_{z \perp}$ fitted by a bilayer tight-binding model~\cite{zhluo23_prl} are shown in Fig.~\ref{fig:sr}. We find that $t_{z \perp}$, which in general defines the splitting  between the bonding and antibonding bands of the two $\dz$ orbitals, increases as system is electron doped. The ratio $t_{z\perp}$/$t_{1x}$ changes from approximately 1.7 to about 2.5 from pristine $\lno$ to $\mathrm{La_2HfNi_2O_7}$. This is in contrast to the hole-doped $\mathrm{La_2SrNi_2O_7}$, where $t_{z\perp}/t_{1x}$ is reduced to $ \sim 1.44$.


Using the constrained random-phase approximation (cRPA)~\cite{aryasetiawan2004frequency}, we also calculate Hubbard interaction parameters $U$ and  Hund's coupling $J_H$ for different compounds. The results are summarized in Table~\ref{tab:JH}, where one can see that while the average
Hubbard $\bar{U}$ is reduced upon electron doping, the relative strength of  Hund's coupling $J_H/U$ increases.  The enhanced $J_H/U$ ratio may boost the Hund's coupling induced SC in the electron doped $\lno$~\cite{cjwu24_hund,mo2025intertwined, ji2025strong}. The cRPA interaction parameters for different substrates can also be found in the Appendix (Table~\ref{tab:UJ}).



\begin{table}
  \caption{Interaction parameters from cRPA. Here thin films with 0.5 UC thickness is used.}
\setlength{\tabcolsep}{2pt}
\setlength{\extrarowheight}{3pt}
\label{tab:JH}
  \centering
  \begin{tabular}{ccccc}
    \hline
      & $\mathrm{La_2ThNi_2O_7}$ & $\mathrm{La_2HfNi_2O_7}$ & $\lno$ & $\mathrm{La_2SrNi_2O_7}$ \\
\hline
    $J_H$ (eV) & 0.62 & 0.60 & 0.58 & 0.49 \\
    $J_H/U$ & 0.22 & 0.22 & 0.15 & 0.19 \\
    $\bar{U}$ & 2.85 & 2.76 & 3.77 & 2.64 \\
    \hline
  \end{tabular}
\end{table}

\subsection{Fermi surface}

\begin{figure}[t!]
\includegraphics[scale=0.40]{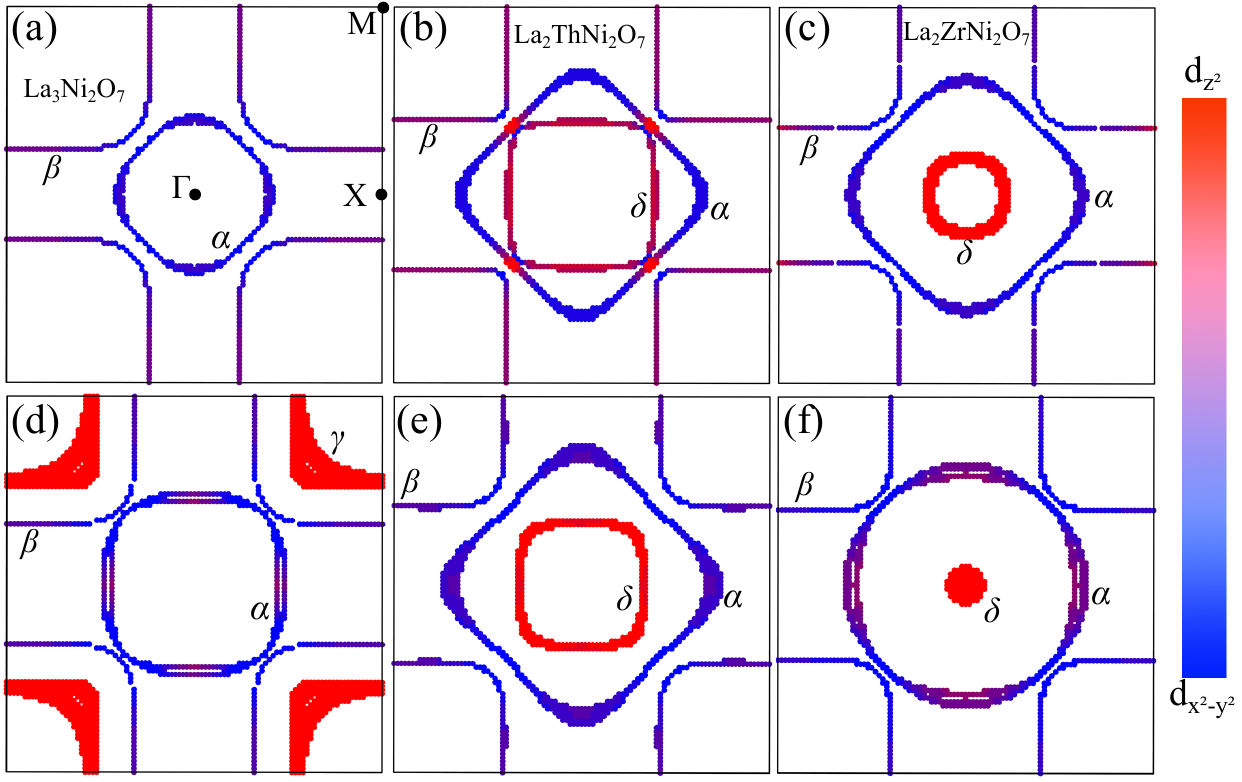}
\caption{Calculated Fermi surfaces of 1-UC $\mathrm{La_{2-x}R_xNi_2O_7}$ thin films. (a--c): Results of $\lno$, $\mathrm{La_2ThNi_2O_7}$, $\mathrm{La_2ZrNi_2O_7}$ on a compressive substrate with $a=3.77$\,\AA; (d--f): Results of  same compounds  but on a tensile substrate with $a=3.905$\,\AA. The color scale indicates the $e_g$ orbital weight.}
\label{fig:fermi-eg}
\end{figure}

Finally,  Fig.~\ref{fig:fermi-eg} displays the Fermi surfaces of $\mathrm{La_{2-x}R_xNi_2O_7}$ under compressive and tensile strains, with blue (red) color scales indicating the $\dx$ ($\dz$) orbital weight. For pristine $\lno$ under $-1.6\%$ compressive strain (Fig.~\ref{fig:fermi-eg}a,  $a=3.77$), the system retains two main pockets ($\alpha$ and $\beta$), consistent with the unstrained configuration and with ARPES measurements~\cite{zxshen25_arpes}. Under $+1.9\%$ tensile strain (Fig.~\ref{fig:fermi-eg}d, $a=3.905$), the anti-bonding band of $\dz$ orbital shifts to higher energy, the hole-like $\gamma$ pocket around M point emerges. 

Upon electron doping (Fig.~\ref{fig:fermi-eg}b--c, Fig.~\ref{fig:fermi-eg}e--f ), the $\alpha$ pocket is significantly enlarged, while the $\beta$ pocket shrinks. For all the four different cases, the $\gamma$ pocket around M points vanishes.  As a result of the shifting down of  anti-bonding band of  $\dz$ orbital (see also~Fig.\ref{fig:band}), a new $\delta$ pocket emerges around $\Gamma$ point,  and it gets enlarged as  substrate changes type from tensile ( $a=3.905$) to compressive ($a=3.77$).

\section{Discussion and Conclusion}
Recent numerical calculations suggest that two distinct pairing mechanisms, namely, Hund's coupling‑induced SC and two‑component SC, can dominate different parameter regimes of the bilayer two‑orbital Hubbard model for $\lno$~\cite{mo2025intertwined}. On the hole‑doped side, when the $\gamma$ band crosses the Fermi level, the enhanced low‑energy DOS of the $\dz$ orbital favors two‑component SC. In contrast, when electron doping or other factors push the $\gamma$ band below the Fermi level, the low-energy degrees of freedom of $\dz$ electrons is suppressed. In this case, the more localized $\dz$ electrons can transfer their strong interlayer antiferromagnetic correlations to $\dx$ electrons via Hund's coupling, driving another SC instability. This picture appears to provide a unified scenario for explaining superconductivity in $\lno$ systems, both with and without a $\gamma$ pocket.

In practice, however, only a limited number of $\lno$ variants have been synthesized. Hence a wide region of the theoretical phase diagram~\cite{mo2025intertwined} remains unexplored. Moreover, experimental uncertainties, such as difficulties in precisely determining the $\gamma$‑band position using angle‑resolved photoemission spectroscopy (ARPES), make it challenging to test the theoretical picture experimentally. Therefore, in addition to hole‑doped $\lno$ (via Sr doping or oxygen content adjustment), the synthesis of electron‑doped $\lno$ is particularly valuable for understanding the pairing mechanism in $\lno$.

In this work, we have performed systematic DFT calculations on $\mathrm{La_{3-x}R_xNi_2O_7}$ (R=Ce, Zr, Hf, Th) to identify viable electron‑doping candiates. Our results establish Zr, Hf, and Th as effective electron dopants, whereas Ce, despite being widely used for electron doping in cuprates, is less efficient in doping electrons to $\lno$ compared to Zr, Hf, and Th. We have  further shown that  these dopants induce an uneven distribution of electron carriers between the $\dx$ and $\dz$ orbitals, which may give rise to distinct pairing mechanisms in the electron-doped regime.   The enhancing of inter-layer $\dz$ hopping amplitute $t_{z \perp}$  may also enhance SC in these systems~\cite{yang2026filling}. In summary, our first-principles calculations fill a critical gap in the exploration of electron-doped $\lno$, provide theoretical guidance for experimental realization, and may contribute to a more unified understanding of the superconducting  $T_c$ in $\lno$ material family.

\textbf{Note added:} While finalizing our work, a  related  work appears~\cite{liu2026enhanced}, which uses model calculations designed for electron-doped $\lno$ to reveal enhanced $s_\pm-$ wave superconductivity. Different from our proposal, Ref.~\cite{liu2026enhanced} suggests to use $\mathrm{La_3Ni_2O_7/La_3Al_2O_7}$ heterostructure to realize electron doping.

\section{Acknowledgment} 
W.W thanks Yi-feng Yang, Dao-xin Yao, and Meng Wang for discussions. This work is supported by the National Natural Science Foundation of China (Grants  No.12494594, No.12274472). We also thank the support from the Research Center for Magnetoelectric Physics of Guangdong Province (Grants No. 2024B0303390001), Guangdong Provincial Quantum Science Strategic Initiative (Grant No. GDZX2401010) and the Fundamental and Interdisciplinary Disciplines Breakthrough Plan of the Ministry of Education of China (JYB2025XDXM403).

\bibliographystyle{apsrev4-2}
\bibliography{ni7}

\onecolumngrid

\section{Appendix}

\twocolumngrid

Here we present supplementary results of $\lno$ thin films on different substrate lattice constants ($a = 3.77$, 3.855, and 3.905\,\AA) and of different dopant compositions. Our results include band structures, density of states, electron-counting data, additional cRPA interaction parameters, and an analysis of the Ce $4f$-electron treatment and the lattice-constant dependence of Ce doping efficacy.

\begin{figure}[t!]
\includegraphics[scale=0.40]{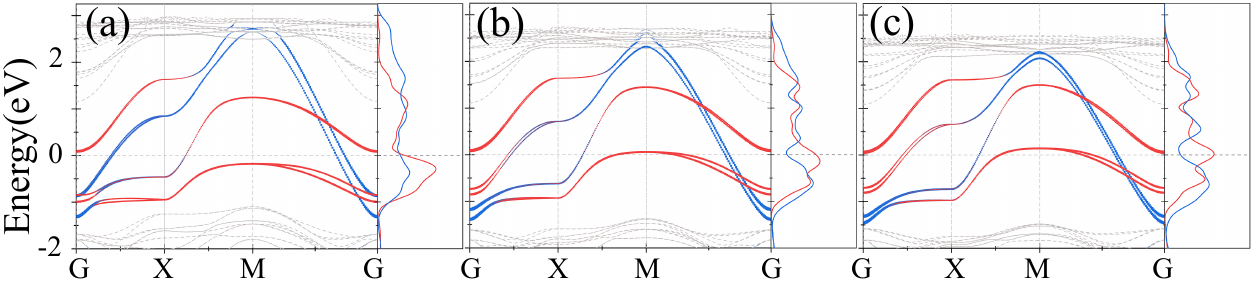}
\caption{(a--c) Band structure and density of states of $\lno$ under in-plane lattice constants $a = 3.77$\,\AA, $3.855$\,\AA, and $3.905$\,\AA.}
\label{fig:band-strain}
\end{figure}

\begin{figure}[t!]
\includegraphics[scale=0.35]{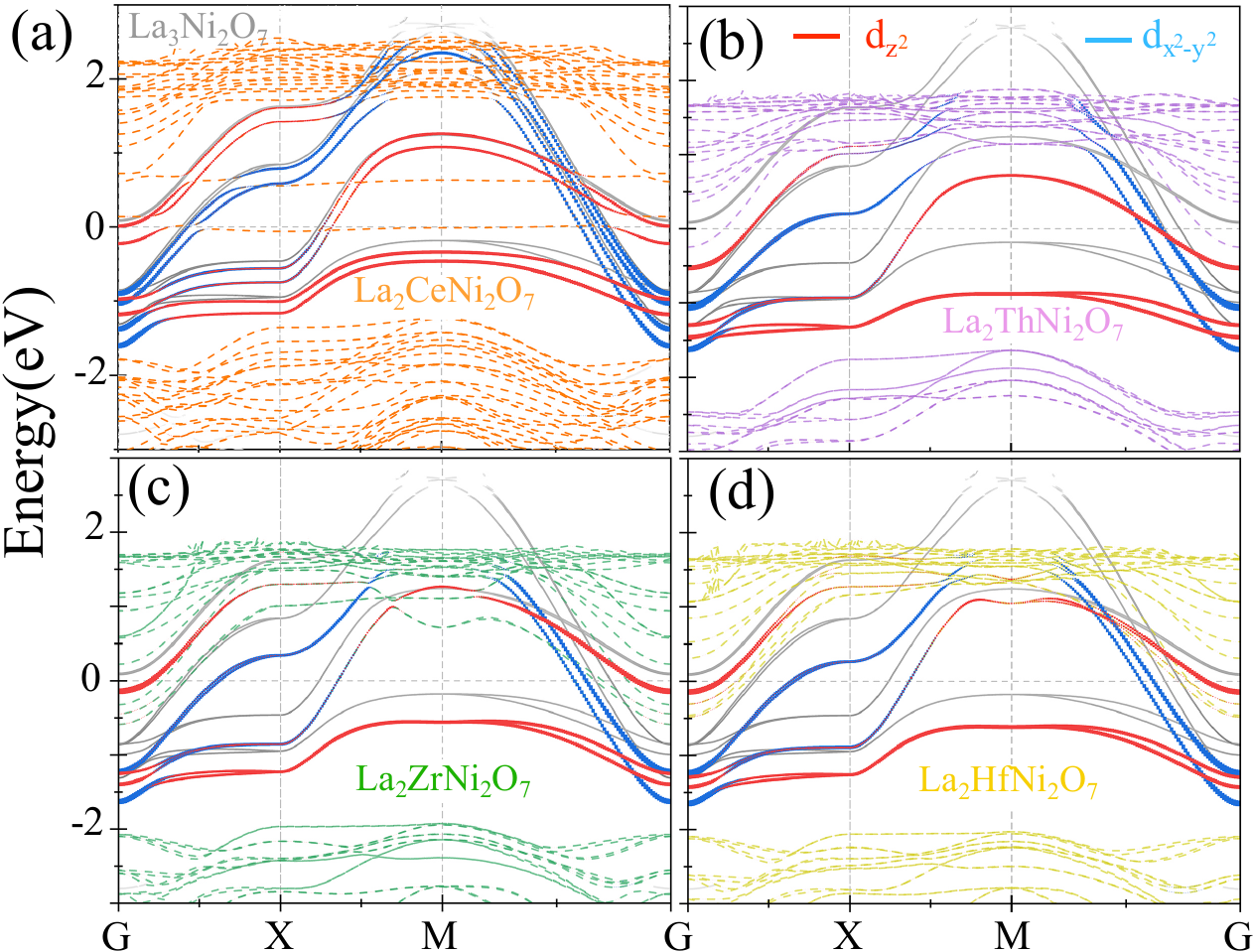}
\caption{Energy bands of 1UC $\mathrm{La_{2-x}R_xNi_2O_7}$ thin films on a substrate with $a=3.77$\,\AA. (a) $\mathrm{La_2CeNi_2O_7}$, (b) $\mathrm{La_2ThNi_2O_7}$, (c) $\mathrm{La_2ZrNi_2O_7}$, (d) $\mathrm{La_2HfNi_2O_7}$. Low-energy bands of pristine $\lno$ are shown by gray lines for reference. For $\mathrm{La_2CeNi_2O_7}$, DFT+$U$ is also applied to the Ce-$4f$ orbitals.}
\label{fig:band-3.77}
\end{figure}

\subsection{Electronic Band Structure and Electron counts under Different Substrate Strains}
Fig.~\ref{fig:band-strain} shows the band structure and DOS of thin-film $\lno$ under three substrate lattice constants. As stain changes from compressive to tensile, the $\dz$ orbital shifts upward and crosses $E_F$, forming a $\gamma$ pocket. 

Fig.~\ref{fig:band-3.77} shows the band structure of 1UC $\mathrm{La_{2-x}R_xNi_2O_7}$ under compressive substrate strain ($a = 3.77$\,\AA). Under compression, the $\gamma$ band lies below $E_F$; additional electron doping shifts the bands further downward, consistent with the electron-count trends in Table~\ref{tab:label}.
Table~\ref{tab:label-3.855} and ~\ref{tab:label-3.905} show the changes in the number of electrons of 1UC $\mathrm{La_{2-x}R_xNi_2O_7}$ under tensile strain.

\subsection{Interaction Parameters under Different Substrate Strains}
Table~\ref{tab:UJ} lists the effective interaction matrix computed by cRPA~\cite{aryasetiawan2004frequency} using the projector method in the Wannier basis,
for two substrates untouched in the main text. The Hund's coupling $J_H$ is found largely insensitive to substrate strain. The ratio $J_H/U$ is also nearly independent of the lattice constant of the substrate.
\begin{table}
  \caption{Interaction parameters (eV)}
\setlength{\tabcolsep}{3pt}
\setlength{\extrarowheight}{3pt}
\label{tab:UJ}
  \centering
  \begin{tabular}{ccccc}
    \hline
      & $\mathrm{La_2ThNi_2O_7}$ & $\mathrm{La_2HfNi_2O_7}$ & $\lno$ & $\mathrm{La_2SrNi_2O_7}$ \\
\hline
   &  & 3.77 Å&  &  \\
\hline
    $J_H$ & 0.62 & 0.60 & 0.58 & 0.49 \\
    $J_H/U$ & 0.22 & 0.22 & 0.15 & 0.19 \\
    $U_x$ & 2.93 & 2.95 & 3.91 & 2.79 \\
    $U_z$ & 2.78 & 2.58 & 3.62 & 2.49 \\
    $\bar{U}$ & 2.85 & 2.76 & 3.77 & 2.64 \\
    \hline
    &  & 3.855 Å& &    \\
\hline
    $J_H$ & 0.62 & 0.60 & 0.59 & 0.50 \\
    $J_H/U$ & 0.23 & 0.22 & 0.16 & 0.20 \\
    $U_x$ & 2.88 & 2.95 & 3.93 & 2.82 \\
    $U_z$ & 2.54 & 2.45 & 3.50 & 2.33 \\
    $\bar{U}$ & 2.71 & 2.70 & 3.71 & 2.53 \\
\hline
  \end{tabular}
\end{table}

\subsection{Treatment of the Ce $4f$ Electrons}

\begin{figure}[t!]
\includegraphics[scale=0.40]{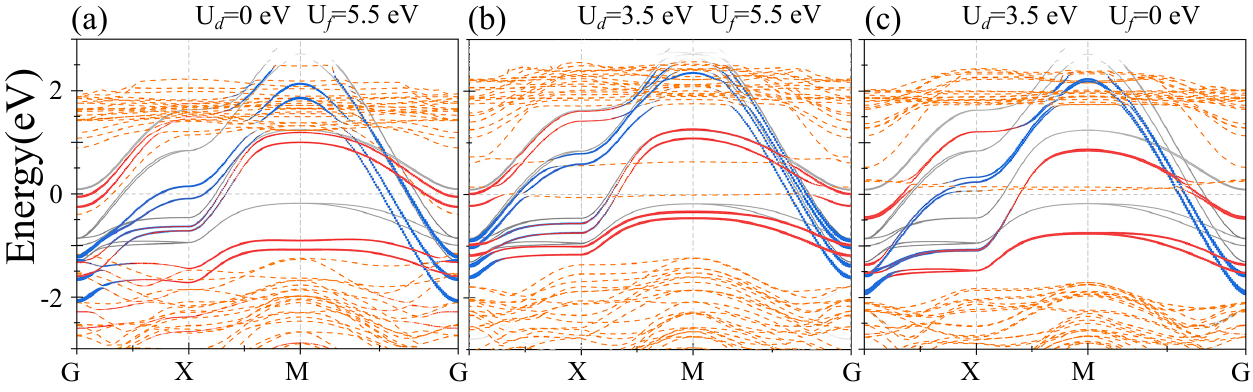}
\caption{(a--c) Band structures of $\mathrm{La_2CeNi_2O_7}$ under compressive substrate strain ($a=3.77$\,\AA) for ($U_d$, $U_f$) = (0, 5.5)\,eV, (3.5, 5.5)\,eV, and (3.5, 0)\,eV, respectively. $U_d$ and $U_f$ are applied to Ni-$3d$ and Ce-$4f$, respectively.}
\label{fig:uf}
\end{figure}

We have employed the standard Ce pseudopotential with DFT+$U$ to describe the strong on-site Coulomb interaction among Ce $4f$ electrons~\cite{been2021electronic}. Fig.~\ref{fig:uf} compares the band structure of $\mathrm{La_2CeNi_2O_7}$ on the $a = 3.77$\,\AA\ substrate for three combinations of $U_d$ (on Ni-$3d$) and $U_f$ (on Ce-$4f$). When both $U_d$ and $U_f$ are present, the band shift is minor and only weak electron doping occurs. Only when one of $U_d$ or $U_f$ is turned off by hand, the low-energy bands can shift substantially downward and introduce electron doping into the system.

\begin{figure}[t!]
\includegraphics[scale=0.38]{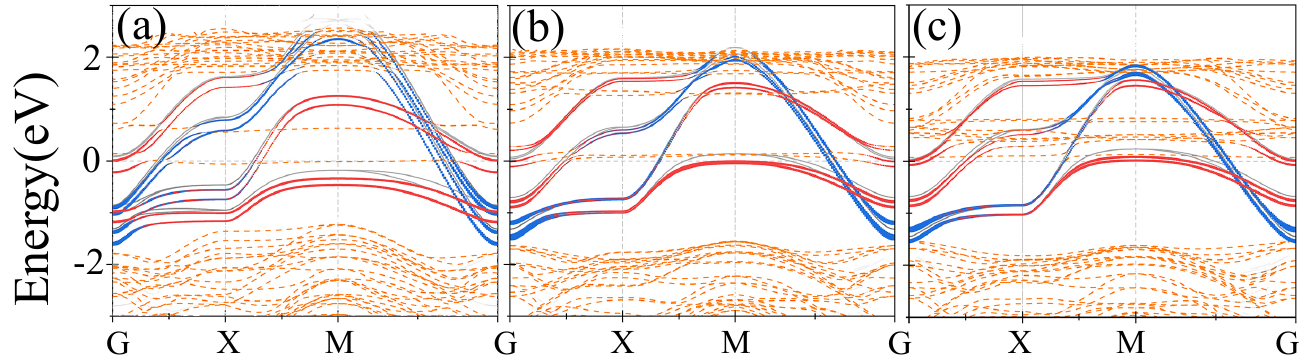}
\caption{(a--c) Band structures of $\mathrm{La_2CeNi_2O_7}$ thin films under in-plane lattice constants of 3.77, 3.905, and 3.975\,\AA, with $U_f = 5.5$\,eV.}
\label{fig:strain-f}
\end{figure}

The limited effectiveness of Ce doping in $\lno$ stands in sharp contrast to its behavior in cuprates. For cuprate thin films~\cite{krockenberger2012universal}, the in-plane lattice constant of a fully relaxed $\mathrm{Nd_2CuO_4}$ film on $\mathrm{SrTiO_3}$ is 3.940\,\AA. Upon Ce doping, the in-plane lattice constant of $\mathrm{Nd_{1.85}Ce_{0.15}CuO_4}$ becomes 3.954\,\AA, while the $c$-axis lattice parameter decreases from 12.126 to 12.082\,\AA. The substitution of $\mathrm{Nd^{3+}}$ ($r = 1.109$\,\AA) by $\mathrm{Ce^{4+}}$ ($r = 0.97$\,\AA) elongates the Cu--O bond, thereby doping the $\mathrm{CuO_2}$ planes. 

To examine the influence of lattice constant on Ce doping in $\lno$, we performed DFT+$U$ calculations on $\mathrm{La_2CeNi_2O_7}$ thin films at $a = 3.77$, 3.905, and 3.975\,\AA. As shown in Fig.~\ref{fig:strain-f}, the magnitude of the Ce-induced band shift is similarly modest at all three lattice constants, which again points to the fact that Ce is generally a weak electron dopant in $\lno$. However, it is worth noting that when the $\gamma$ band is very flat and near Fermi level, a mild shift of bands  may cause significant carrier doping. In  Fig.~\ref{fig:strain-f},  for the pristine $\lno$, under compression ($a=3.77$\,\AA) the $\gamma$ band already lies below $E_F$, whereas under tension ($a=3.905$, $3.975$\,\AA) it is  just above $E_F$.  For the latter two cases, the modest Ce-induced shift can bring the near flat  $\gamma$ band down into Fermi sea, leading to a significant increment of the $\dz$ electron count. Thus, while Ce is generally a weak electron dopant, its doping effect is not negligible on substrates when $\gamma$-band is flat and accidentally lies just above $E_F$, see also Table~\ref{tab:label-3.855} and Table~\ref{tab:label-3.905}.

\begin{figure}[t!]
\includegraphics[scale=0.40]{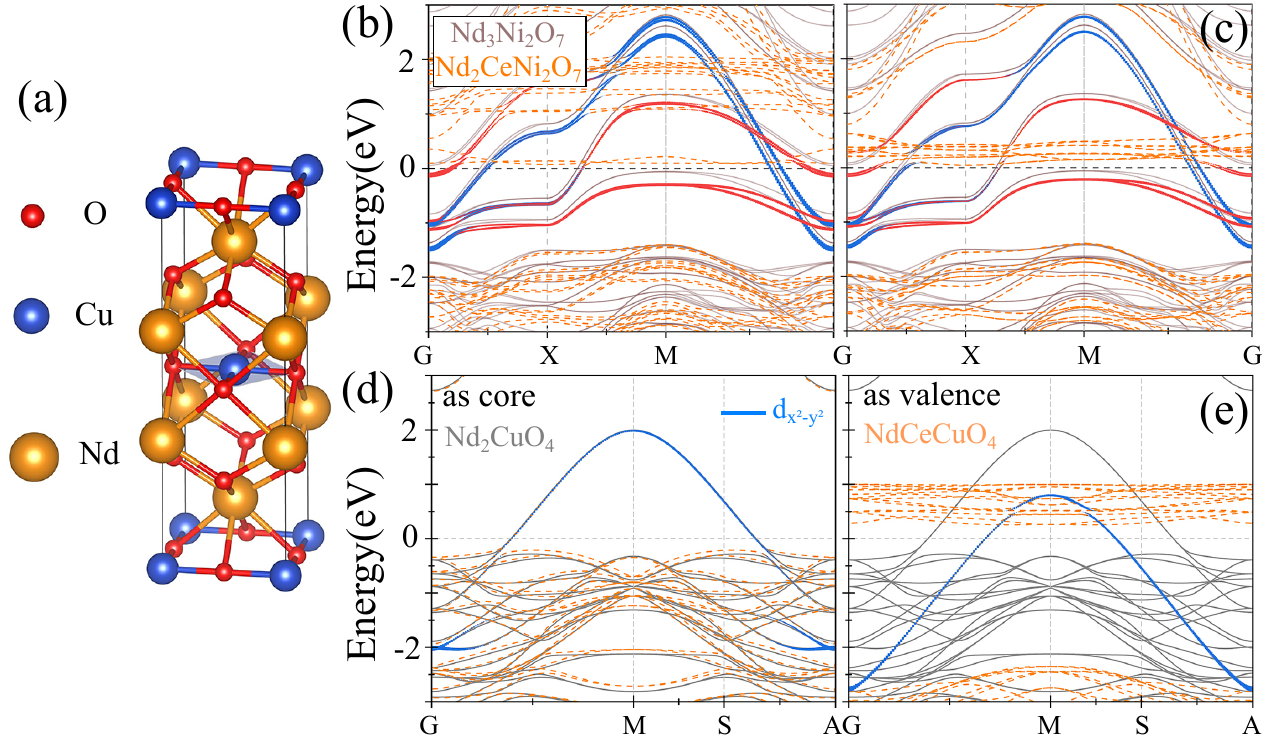}
\caption{(a) Crystal structure of $\mathrm{Nd_2CuO_4}$. (b)--(c) Band structures of $\mathrm{Nd_2CeNi_2O_7}$ with $U_f = 0$ and $U_f = 5.5$\,eV, respectively. (d) Band structure of $\mathrm{NdCeCuO_4}$ with Ce-$4f$ treated as core states. (e) Same as (d) with Ce-$4f$ treated as valence electrons and $U_f = 5.5$\,eV.}
\label{fig:cu-dope}
\end{figure}

We further examine Nd-based nickelates and compare with NCCO. Nd substitution ($r_{\mathrm{Nd^{3+}}} = 1.109$\,\AA, smaller than La$^{3+}$) mimics lattice compression. Figs.~\ref{fig:cu-dope}(b) and (c) show the band structures of $\mathrm{Nd_2CeNi_2O_7}$ with $U_f = 0$ and $U_f = 5.5$\,eV: the doping effect appears only when $U_f$ is turned off, mirroring the La-based results in Fig.~\ref{fig:uf}. For NCCO, treating Ce-$4f$ as core states yields negligible doping (Fig.~\ref{fig:cu-dope}(d)), while promoting $4f$ to the valence restores the pronounced downward shift (Fig.~\ref{fig:cu-dope}(e)), confirming that the $4f$ valence treatment is essential for describing electron doping in cuprates.

\begin{table*}
  \caption{Electron counts for individual Ni-$d$ orbitals, $t_{2g}$ orbitals, $e_g$ orbitals, as well as in-plane and out-of-plane components with a=3.855 Å}
\setlength{\tabcolsep}{8pt}
\setlength{\extrarowheight}{3pt}
\label{tab:label-3.855}
  \centering
  \begin{tabular}{cccccccccc}
    \hline
      & Ni-$d$ & $t_{2g}$ & $\dz$ & $\dx$ & In & Out & $\Delta$In & $\Delta$Out \\
\hline
    $\lno$ & 8.338 & 5.812 & 1.288 & 1.237 & 3.336 & 2.868 & -  & - \\
    \hline
    $\mathrm{La_{2.5}Ce_{0.5}Ni_2O_7}$ & 8.376 & 5.793 & 1.380 & 1.203 & 3.326 & 2.981 & -1.0\% & 11.3\% \\
    $\mathrm{La_2CeNi_2O_7}$ & 8.398 & 5.775 & 1.403 & 1.220 & 3.364 & 3.020 & 2.8\% & 15.2\% \\
    \hline
    $\mathrm{La_{2.75}Zr_{0.25}Ni_2O_7}$ & 8.357 & 5.788 & 1.398 & 1.172 & 3.292 & 2.997 & -4.4\% & 12.8\% \\
    $\mathrm{La_{2.5}Zr_{0.5}Ni_2O_7}$ & 8.380 & 5.772 & 1.373 & 1.235 & 3.382 & 2.983 & 4.6\% & 11.5\% \\
    $\mathrm{La_2ZrNi_2O_7}$ & 8.421 & 5.736 & 1.385 & 1.301 & 3.486 & 3.010 & 15.2\% & 14.1\% \\
    \hline
    $\mathrm{La_{2.75}Hf_{0.25}Ni_2O_7}$ & 8.368 & 5.795 & 1.108 & 1.165 & 3.283 & 3.006 & -5.3\% & 13.8\% \\
    $\mathrm{La_{2.5}Hf_{0.5}Ni_2O_7}$ & 8.399 & 5.774 & 1.391 & 1.234 & 3.387 & 3.003 & 5.1\% & 13.5\% \\
    $\mathrm{La_2HfNi_2O_7}$ & 8.447 & 5.733 & 1.396 & 1.318 & 3.516 & 3.021 & 18.0\% & 15.3\% \\
    \hline
    $\mathrm{La_{2.75}Th_{0.25}Ni_2O_7}$ & 8.373 & 5.797 & 1.379 & 1.197 & 3.316 & 2.980 & -2.0\% & 11.2\% \\
    $\mathrm{La_{2.5}Th_{0.5}Ni_2O_7}$ & 8.395 & 5.777 & 1.430 & 1.188 & 3.327 & 3.045 & -0.9\% & 17.7\% \\
    $\mathrm{La_2ThNi_2O_7}$ & 8.459 & 5.735 & 1.478 & 1.246 & 3.430 & 3.115 & 9.4\% & 24.7\% \\
    \hline
    $\mathrm{La_{2.75}Sr_{0.25}Ni_2O_7}$ & 8.309 & 5.824 & 1.252 & 1.233 & 3.309 & 2.813 & -2.7\% & -5.5\% \\
    $\mathrm{La_{2.5}Sr_{0.5}Ni_2O_7}$ & 8.291 & 5.828 & 1.240 & 1.224 & 3.279 & 2.783 & -5.7\% & -8.5\% \\
    $\mathrm{La_2SrNi_2O_7}$ & 8.239 & 5.846 & 1.168 & 1.224 & 3.232 & 2.673 & -10.3\% & -19.5\% \\
    \hline
    $\Delta$In=($\mathrm{In_R}$ - $\mathrm{In_{La}}$) $\times$ 100\% \\
  \end{tabular}
\end{table*}

\begin{table*}
  \caption{Electron counts for individual Ni-$d$ orbitals, $t_{2g}$ orbitals, $e_g$ orbitals, as well as in-plane and out-of-plane components with a=3.905 Å}
\setlength{\tabcolsep}{8pt}
\setlength{\extrarowheight}{3pt}
\label{tab:label-3.905}
  \centering
  \begin{tabular}{cccccccccc}
    \hline
      & Ni-$d$ & $t_{2g}$ & $\dz$ & $\dx$ & In & Out & $\Delta$In & $\Delta$Out \\
\hline
    $\lno$ & 8.327 & 5.807 & 1.231 & 1.289 & 3.381 & 2.806 & -  & - \\
    \hline
    $\mathrm{La_{2.5}Ce_{0.5}Ni_2O_7}$ & 8.308 & 5.790 & 1.265 & 1.253 & 3.362 & 2.852 & -1.9\% & 4.6\% \\
    $\mathrm{La_2CeNi_2O_7}$ & 8.424 & 5.808 & 1.454 & 1.162 & 3.298 & 3.063 & -8.3\% & 25.6\% \\
    \hline
    $\mathrm{La_{2.75}Zr_{0.25}Ni_2O_7}$ & 8.357 & 5.782 & 1.336 & 1.239 & 3.356 & 2.929 & -2.4\% & 12.2\% \\
    $\mathrm{La_{2.5}Zr_{0.5}Ni_2O_7}$ & 8.364 & 5.773 & 1.307 & 1.285 & 3.422 & 2.910 & 4.1\% & 10.4\% \\
    $\mathrm{La_2ZrNi_2O_7}$ & 8.401 & 5.741 & 1.313 & 1.347 & 3.523 & 2.933 & 14.2\% & 12.7\% \\
    \hline
    $\mathrm{La_{2.75}Hf_{0.25}Ni_2O_7}$ & 8.359 & 5.788 & 1.355 & 1.216 & 3.329 & 2.947 & -5.2\% & 14.0\% \\
    $\mathrm{La_{2.5}Hf_{0.5}Ni_2O_7}$ & 8.386 & 5.774 & 1.332 & 1.280 & 3.424 & 2.938 & 4.3\% & 13.2\% \\
    $\mathrm{La_2HfNi_2O_7}$ & 8.422 & 5.737 & 1.329 & 1.356 & 3.544 & 2.950 & 16.3\% & 14.4\% \\
    \hline
    $\mathrm{La_{2.75}Th_{0.25}Ni_2O_7}$ & 8.355 & 5.791 & 1.307 & 1.256 & 3.370 & 2.900 & -1.1\% & 9.4\% \\
    $\mathrm{La_{2.5}Th_{0.5}Ni_2O_7}$ & 8.384 & 5.773 & 1.381 & 1.230 & 3.363 & 2.990 & -1.8\% & 18.4\% \\
    $\mathrm{La_2ThNi_2O_7}$ & 8.449 & 5.731 & 1.445 & 1.273 & 3.450 & 3.078 & 6.9\% & 27.1\% \\
    \hline
    $\mathrm{La_{2.75}Sr_{0.25}Ni_2O_7}$ & 8.300 & 5.819 & 1.195 & 1.286 & 3.355 & 2.751 & -2.6\% & -5.5\% \\
    $\mathrm{La_{2.5}Sr_{0.5}Ni_2O_7}$ & 8.281 & 5.825 & 1.173 & 1.283 & 3.330 & 2.712 & -5.1\% & -9.4\% \\
    $\mathrm{La_2SrNi_2O_7}$ & 8.223 & 5.842 & 1.124 & 1.283 & 3.253 & 2.626 & -12.8\% & -18.0\% \\
    \hline
    $\Delta$In=($\mathrm{In_R}$ - $\mathrm{In_{La}}$) $\times$ 100\% \\
  \end{tabular}
\end{table*}

\end{document}